\documentclass[showpacs,preprintnumbers,nofootinbib,12pt]{revtex4-1}

\usepackage{graphicx}  %commented out by MK
\usepackage{dcolumn}
\usepackage{bm}
\usepackage{amsmath,amssymb,amsfonts}
\usepackage{latexsym}
\usepackage{color}
\def\nn    {\nonumber}
\def\bhpm{$bH^\pm~$}

\def\fbi{fb$^{-1}$\;}

\begin{document}

\title{\boldmath  Implications of the heavy Higgs induced single-top, same-sign top and triple-top productions at the LHC}

\author{Tanmoy Modak}
\affiliation{Institut f{\"u}r Theoretische Physik, Universit{\"a}t Heidelberg, 69120 Heidelberg, Germany}
\bigskip

% \date{\today}

\begin{abstract}
In this brief review we study the discovery potential of heavy Higgs induced single-top, same-sign top and triple-top quark productions at the 
Large Hadron Collider (LHC). The context is general two Higgs doublet model where such processes can be induced by additional 
Yukawa couplings and sub-Tev Higgs bosons. We show that these processes can be discovered at the ongoing or future high luminosity
LHC run. Discovery of these multi-top productions may shed light on the mechanism behind baryogenesis or cosmic inflation.

\end{abstract}

\maketitle

%-----------------------------------------------------------------------------------------------------------------------------------
%	Introduction
\section{Introduction}
%-----------------------------------------------------------------------------------------------------------------------------------
The study of top quark productions is a staple program at the LHC.
As a proton-proton collider LHC produces top-quarks abundantly and provides unique window 
to study multi-top quark productions. In the early part of Run-1 ATLAS and CMS collaborations swiftly rediscovered 
$t\bar t$ production~\cite{ATLAS:2010zaw,CMS:2010uxk}.  Subsequently LHC also measured the single-top productions both
in the $t$-~\cite{ATLAS:2012byx,CMS:2012xhh,ATLAS:2016qhd,CMS:2016lel} and
$tW$~\cite{CMS:2014fut,ATLAS:2015igu,ATLAS:2016ofl,CMS:2021vqm} channels. Searches for the subdominant $s$-channel single-top production 
are also performed by both the collaborations~\cite{ATLAS:2015jmq,CMS:2016xoq,ATLAS:2022wfk}. On the other hand, the triple-top production is at merely few fb in 
the Standard Model (SM)~\cite{Barger:2010uw} but the QCD initiated four-top production is at ${\cal O}(10)$~fb. While none of the experiments have covered the SM triple-top 
in their search programs so far however the four-top productions are searched by ATLAS~\cite{ATLAS:2018kxv,ATLAS:2020hrf}
and CMS~\cite{CMS:2017ocm,CMS:2019rvj}.

Multi-top productions are also possible in the minimal extensions of the SM such as the two Higgs doublet model (2HDM)~\cite{Djouadi:2005gj,Branco:2011iw}.
After the discovery of the 125 GeV Higgs boson $h$~\cite{h125_discovery} presence of additional Higgs doublet seems more plausible. In such 
model in addition to the observed $h$ boson heavy additional scalars $A$ and $H$ or the charged Higgs boson $H^\pm$ may also induce different 
top quark productions. Indeed, extensive efforts have been directed towards these heavy Higgs induced multi-top quark productions e.g. 
$pp\to A/H \to t \bar t$~\cite{Gaemers:1984sj,Dicus:1994bm,Bernreuther:1997gs,Frederix:2007gi,Carena:2016npr,BuarqueFranzosi:2017jrj,ATLAS:2017snw,CMS:2019pzc,Aad:2020klt},
$pp\to t \bar t A/H \to t \bar t  t \bar t$~\cite{Dicus:1994bm,Craig:2015jba,Kanemura:2015nza,Gori:2016zto,Craig:2016ygr,CMS:2019rvj},
and $pp\to \bar t {H}^+ (b) \to t \bar t \bar b (b)$~\cite{Plehn:2002vy,Gunion:2002zf,Boos:2003yi,Akeroyd:2016ymd,Craig:2012vn,
Carena:2013qia,Carena:2013ooa,LHCHiggsCrossSectionWorkingGroup:2016ypw,Harlander:2011aa,CMS:2015lsf,ATLAS:2015nkq,ATLAS:2018ntn,CMS:2019rlz,ATLAS:2020jqj,CMS:2020imj,ATLAS:2021upq} at the LHC.

In this brief review we focus on some additional multi-top productions 
in the context of general 2HDM (g2HDM). In the absence of any discrete 
symmetry both the doublets in g2HDM couple with up- and down-type
fermions. After diagonalization of the fermion mass matrices two different
Yukawa couplings $\lambda_{ij}^F =({\sqrt{2}m_i^F}/{v})\, \delta_{ij}$  
(with $v \simeq 246$ GeV) and $\rho_{ij}^F$ emerge with
$F$ denoting up- and down-type quarks and charged-leptons. While,
the $\lambda_{ij}^F$  matrices are real and diagonal but the $\rho_{ij}^F$
are in general complex and nondiagonal.

It has been shown that successful electroweak baryogenesis (EWBG) is possible for sizeable complex 
$\rho_{\tau\tau}$~\cite{Chiang:2016vgf}, $\rho_{tt}$ and $\rho_{tc}$~\cite{Fuyuto:2017ewj}~\footnote{Similar mechanisms are also discussed in 
Refs.~\cite{deVries:2017ncy,Cline:2021dkf,Enomoto:2021dkl,Enomoto:2022rrl}.}, 
and $\rho_{bb}$~\cite{Modak:2018csw}. Motivated by this Refs.~\cite{Kohda:2017fkn,Hou:2018zmg} 
(for similar discussion see also Refs.~\cite{Hou:1997pm,Altmannshofer:2016zrn,Iguro:2017ysu,Altmannshofer:2019ogm,Hou:2020ciy,Hou:2020chc})
have analyzed the possibility of discovering $cg\to t A/tH\to t t \bar c$ (denoted as same-sign top)
process at the LHC. The process is induced by $\rho_{tc}$ coupling and can be searched via $pp\to t A/tH +X \to t t \bar c +X$ ($X$ is inclusive activity) followed 
by semileptonic decay of both the top quarks constituting a clean same-sign dilepton signature. Furthermore, if both  $\rho_{tc}$ and  $\rho_{tt}$ are nonvanishing
one may have $cg\to t A/tH\to t t \bar t$ (triple-top)~\cite{Kohda:2017fkn,Hou:2019gpn,ATLAS:2022xpz} and 
$cg\to b H^+ \to b t \bar b$ (single-top) production. At the LHC the former process can be searched 
via  $pp\to t A/tH +X \to t t \bar t +X$ with all three top quarks decaying semileptonically. The charged Higgs induced single-top 
process cane be searched via $pp\to b H^+ +X \to b t \bar b +X$~\cite{Ghosh:2019exx}~\footnote{For $\bar cbH^+$ induced
similar productions see also Refs.~\cite{Diaz-Cruz:2009ysj,Gori:2017tvg,Nierste:2019fbx,Desai:2022zig}.}.

If cosmic inflation is realized in the g2HDM it may favor nearly degenerate $A$, $H$, $H^\pm$ as shown in Refs.~\cite{Lee:2021rzy,Modak:2020fij}.
It has been found that for mass and width degenerate $A$ and $H$ would lead to exact cancellation of the $cg\to t A\to t t \bar c$  
and $cg\to t H\to t t \bar c$ amplitudes~\cite{Kohda:2017fkn}. This would lead to diminished signature for the same-sign top signature while
while no such cancellation exist for $cg\to b H^+ \to b t \bar b$ and $cg\to t A/tH\to t t \bar t$.
Therefore, vanishingly small same-sign signature but discovery of $cg\to b H^+ \to b t \bar b$ and  $cg\to t A/tH\to t t \bar t$  may provide indirect evidence
for inflation.

The paper is organized as follows. 
In Sec.~\ref{sec:param} we discuss in detail the model framework  and relevant constraints
for the parameter space. We first start our discussion with the discovery prospect of the single-top i.e.
$pp\to b H^+ +X \to b t \bar b +X$ signature in Sec.~\ref{sec:bhpm}. The Sec.~\ref{sec:sstop} and Sec.~\ref{sec:3top}
are dedicated for the same-sign top and triple-top signatures respectively.
We close with implications of these discovery and a outlook in Sec.~\ref{disc}.

%%%%%%%%%%%%%%%%%%%%%%%%%%%%%%%%%%%%%%%%%%%%%%%%%%%%%%%%%%%%%%%%%%%%%%%%%%%
\section{ Framework parameter Space}\label{sec:param}
%%%%%%%%%%%%%%%%%%%%%%%%%%%%%%%%%%%%%%%%%%%%%%%%%%%%%%%%%%%%%%%%%%%%%%%%%%%
We consider the most general $CP$-conserving two Higgs doublet potential which is written in the Higgs basis as~\cite{Davidson:2005cw,Hou:2017hiw}

\begin{align}
 V(\Phi,\Phi') &= \mu_{11}^2|\Phi|^2 + \mu_{22}^2|\Phi'|^2
            - (\mu_{12}^2\Phi^\dagger\Phi' + h.c.)  + \frac{\eta_1}{2}|\Phi|^4 + \frac{\eta_2}{2}|\Phi'|^4
           + \eta_3|\Phi|^2|\Phi'|^2  + \eta_4 |\Phi^\dagger\Phi'|^2 \nn \\
 & \quad \quad \quad \quad + \left[\frac{\eta_5}{2}(\Phi^\dagger\Phi')^2
     + \left(\eta_6 |\Phi|^2 + \eta_7|\Phi'|^2\right) \Phi^\dagger\Phi' + h.c.\right],
\label{pot}
\end{align}
with the $\eta_i$s are the quartic couplings, $\mu_{ij}$s are dimension two mass parameters.
The vacuum expectation value $v$ arises from the  $\Phi$ doublet 
via the stationary condition  $\mu_{11}^2=-\frac{1}{2}\eta_1 v^2$, 
while $\left\langle \Phi'\right\rangle =0$ (hence $\mu_{22}^2 > 0$).
The second stationary condition is $\mu_{12}^2 = \frac{1}{2}\eta_6 v^2$ and the 
mixing angle $\gamma$
\begin{align}
 c_\gamma^2 = \frac{\eta_1 v^2 - m_h^2}{m_H^2-m_h^2},~\quad \quad \sin{2\gamma} = \frac{2\eta_6 v^2}{m_H^2-m_h^2}.
\end{align}
diagonalizes the CP-even mass matrix of $h$, $H$, and satisfies~\cite{Davidson:2005cw,Hou:2017hiw} where we used shorthand for $c_\gamma$ for $\cos\gamma$. In the  
alignment limit the $\cos\gamma$ goes to zero limit.
The  masses of the neutral and charged scalars are found as
\begin{align}
&m_{A}^2 = \frac{1}{2}(\eta_3 + \eta_4 - \eta_5) v^2+ \mu_{22}^2,\\
 &m_{h,H}^2 = \frac{1}{2}\bigg[m_A^2 + (\eta_1 + \eta_5) v^2 \mp \sqrt{\left(m_A^2+ (\eta_5 - \eta_1) v^2\right)^2 + 4 \eta_6^2 v^4}\bigg],\\
 &m_{H^+}^2 = \frac{1}{2}\eta_3 v^2+ \mu_{22}^2.
\end{align}

The fermion interactions with the  scalar bosons $h$, $H$, $A$ and $H^+$ is expressed as~\cite{Davidson:2005cw}
\begin{align}
\mathcal{L} = %\supset
-&\frac{1}{\sqrt{2}} \sum_{F = U, D, L}
 \bar F_{i} \bigg[\big(-\lambda^F_{ij} s_\gamma + \rho^F_{ij} c_\gamma\big) h
 +\big(\lambda^F_{ij} c_\gamma + \rho^F_{ij} s_\gamma\big)H -i ~{\rm sgn}(Q_F) \rho^F_{ij} A\bigg]  P_R F_{j}\nn\\
 &-\bar{U}_i\big[(V\rho^D)_{ij} P_R-(\rho^{U\dagger}V)_{ij} P_L\big]D_j H^+ - \bar{\nu}_i\rho^L_{ij} P_R L_j H^+ +{\rm H.c.},\label{eff}
\end{align}
where $V$ is the Cabibbo-Kobayashi-Maskawa (CKM) matrix and $P_{L,R}\equiv (1\mp\gamma_5)/2$ with $i,j = 1, 2, 3$ are generation indices
in flavor space. The $U$, $D$ and $L$ denotes the matrices  
$U=(u,c,t)$, $D = (d,s,b)$, $L=(e,\mu,\tau)$ and $\nu=(\nu_e,\nu_\mu,\nu_\tau)$ whereas 
the matrices $\lambda^F_{ij}\; (\equiv \delta_{ij}\sqrt{2}m_i^F/v)$ and $\rho^F_{ij}$
two Yukawa couplings. The $\lambda^F$ matrices are real and diagonal but $\rho^F$ matrices
are in general in general nondiagonal and complex. The $\rho^F_{ii}$ matrices may follow the same flavor organization principle
as in SM i.e. $\rho^F_{ii}\sim \lambda^F_i$  with suppressed off-diagonal elements which simply means
$\rho^U_{tt}\sim \lambda^U_t$, $\rho^D_{bb}\sim \lambda^D_b$ etc. In what follows we drop the superscript $F$ and set all $\rho_{ij}$ couplings to zero except for $\rho_{tt}$ and 
$\rho_{tc}$ and assume them to be real for simplicity. 

Our focus of interest is to study the discovery potential of the $cg\to b H^+ \to b t \bar b$ (single-top), $cg\to t A/tH\to t t \bar c$ (same-sign top)
and $cg\to t A/tH\to t t \bar t$ (triple-top) productions at the LHC and their respective implications for baryogenesis and inflation.
The first and last processes are induced if both $\rho_{tc}$ and $\rho_{tt}$ are nonvanishing, whereas, the same-sign top is induced by the $\rho_{tc}$ 
coupling alone. We remark here that similar final state topology can be induced by $\rho_{tu}$ which we shall not cover in the current manuscript 
except for the same-sign top i.e. $ug\to t A/tH\to t t \bar u$. Here we assume  $\rho_{tc}$ and $\rho_{tt}$ are real but 
the signatures discussed will remain unchanged if they are complex.
\begin{figure*}[h]
\center
\includegraphics[width=.4 \textwidth]{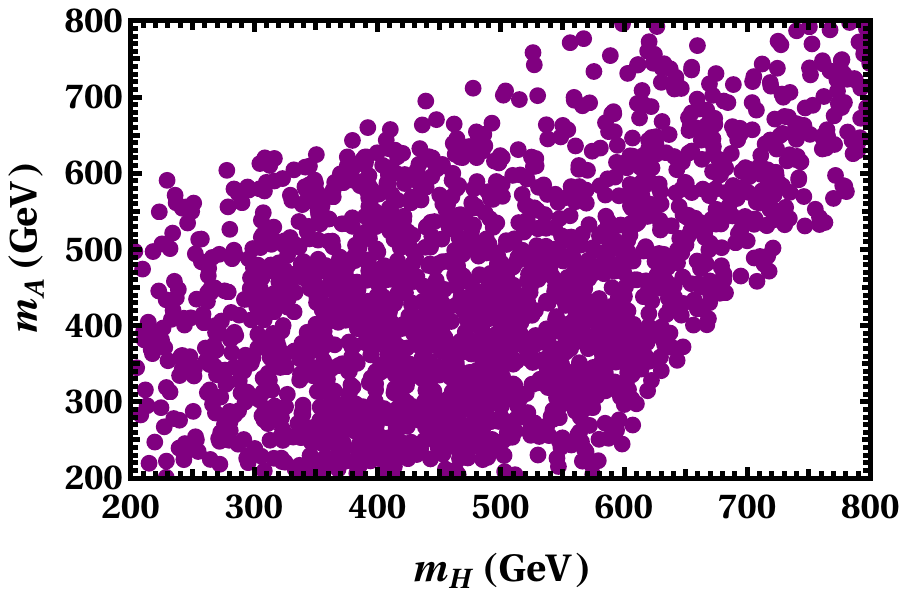}
\includegraphics[width=.4 \textwidth]{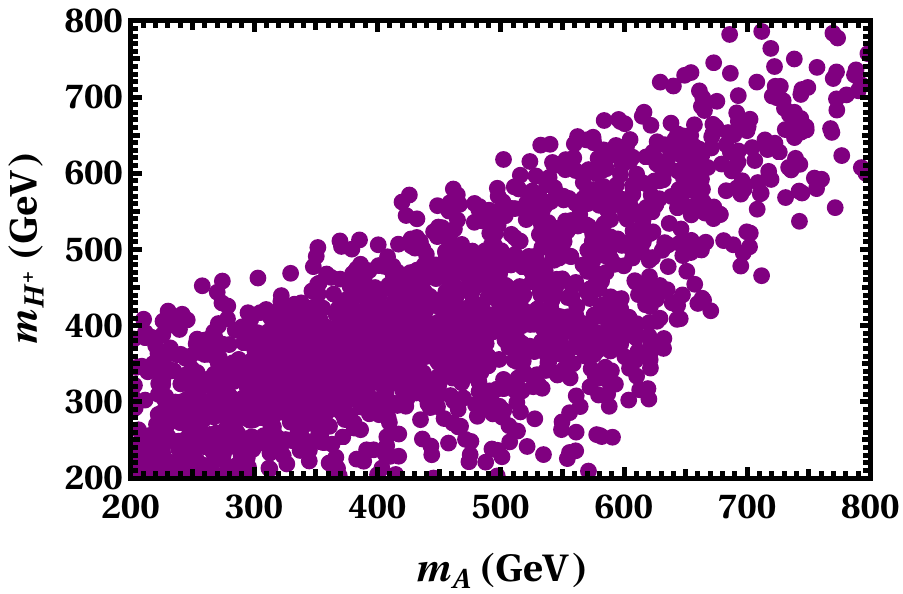}
\caption{The available parameter space that satisfy
after satisfy positivity, perturbativity, tree-level unitarity and EW precision observables are plotted in the $m_A$--$m_H$ and $m_A$--$m_{H^+}$ plane.}
\label{scaned}
\end{figure*}  

Before exploring the discovery prospect let us first discuss the available parameter space.
The dynamical parameters in Eq.~\eqref{pot} need to satisfy
the positivity, perturbativity and tree-level unitarity. We check these constraints
in the public tool 2HDMC~\cite{Eriksson:2009ws} by first expressing the quartic couplings $\eta_1$, $\eta_{3{\rm -}6}$ in terms of 
$m_h^2$, $m_{H^+}^2$, $m_A^2$, $m_H^2$, $\gamma$, $\mu_{22}^2$ and $v$ as~\cite{Davidson:2005cw}
\begin{align}
& \eta_1 = \frac{m_h^2 s_\gamma^2 + m_H^2 c_\gamma^2}{v^2},\\
& \eta_3 =  \frac{2(m_{H^+}^2 - \mu_{22}^2)}{v^2},\\
& \eta_4 = \frac{m_h^2 c_\gamma^2 + m_H^2 s_\gamma^2 -2 m_{H^+}^2+m_A^2}{v^2},\\
& \eta_5 =  \frac{m_H^2 s_\gamma^2 + m_h^2 c_\gamma^2 - m_A^2}{v^2},\\
& \eta_6 =  \frac{(m_h^2 - m_H^2)(-s_\gamma)c_\gamma}{v^2}.
\end{align}
The quartic couplings $\eta_2$ and $\eta_7$ do not enter scalar masses.
We generate the phenomenological parameters 
$\gamma$, $m_A$, $m_H$, $m_{H^+}$, $\mu_{22}$, $\eta_2$, $\eta_7$ randomly
in the following ranges:
$\mu_{22} \in [0, 1000]$\,GeV,
$m_{H^+} \in [200, 800]$\,GeV,
$m_A \in [200, 800]$\,GeV,
$m_H = \in [200, 800]$, 
$\eta_2 \in [0, 5]$, $ \eta_7 \in [-5, 5]$, 
with $m_h = 125$\,GeV and $c_\gamma = 0$ held fixed.
These randomly generated parameters are then fed into 
2HDMC~\cite{Eriksson:2009ws} for scanning in the Higgs basis~\footnote{The input parameters in the Higgs basis in 2HDMC 
are $\Lambda_{1-7}$ and $m_{H^+}$ with $v \simeq 246$ GeV is implicit. In order to match the convention we take $-\pi/2\leq \gamma \leq \pi/2$ and
identify $\Lambda_{1-7}$  as $\eta_{1-7}$.}. We further demand the magnitude of all quartic couplings should be $\leq 5$ conservatively.
Further the constraints from precision electroweak observables~\cite{Peskin:1991sw} further restricts the parameter space~\cite{Froggatt:1991qw,Haber:2015pua} which 
we have included in our analysis. 
The parameter sets that passed the positivity, perturbativity and tree-level unitarity conditions from 2HDMC, 
are also tested for the precision electroweak observables~\cite{Baak:2014ora}. 
These final scanned points are plotted in the  $m_{H}$--$m_A$ and $m_{H^+}$--$m_A$ in Fig.~\ref{scaned} which 
illustrates that finite sub-TeV parameter space still exist.

\begin{figure*}[htbp]
\center
\includegraphics[width=.4 \textwidth]{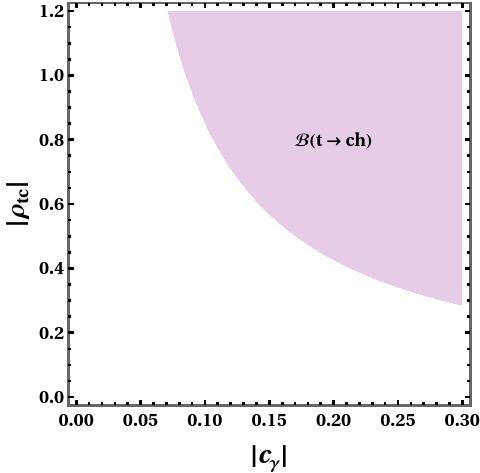}
\caption{
Exclusion limits from $\mathcal{B}(t\to c h)< 9.4\times 10^{-4}$ in the 
for $|\rho_{tc}|$--$c_\gamma$ plane.
}
 \label{tchexclu}
\end{figure*}

We now briefly discuss the constraints on $\rho_{tc}$ and $\rho_{tt}$
There exists several indirect and direct constraints on $\rho_{tc}$.
For nonzero $c_\gamma$ the $\rho_{tc}$ receives significant constraints from
branching ratio of the $t\to c h$~\cite{Hou:1991un}.
The ATLAS~\cite{ATLAS:2022gzn} and CMS~\cite{CMS:2021gfa} both set same 95\% CL upper limit $\mathcal{B}(t\to c h)< 9.4\times 10^{-4}$ with full Run-2 data
which is shown in Fig.~\ref{tchexclu} by purple shaded region.
It is clear from Fig~\ref{tchexclu} that $|\rho_{tc}| \gtrsim 0.8$ is excluded at $95\%$ CL for $c_\gamma \sim 0.1$. The limit becomes stringent
for larger $c_\gamma$. 

Further $\rho_{tc}$ also receives more constraints albeit milder from
$B_{s}$-$\overline{B}_{s}$ mixing and $\mathcal{B}(B\to X_s\gamma)$~\cite{Altunkaynak:2015twa}. 
Such constraints arise through the loop of charm quarks
and a charged Higgs involving $\rho_{tc}$ but reinterpreting the limits from Ref.~\cite{Crivellin:2013wna} one finds
$|\rho_{tc}|\gtrsim 0.9~(1.2)$ is excluded at $2\sigma$ for $m_{H^+}= 300~(500)$ GeV.  
However, the most stringent limit on $\rho_{tc}$ arise from
the CMS search for the SM four-top production~\cite{CMS:2019rvj}, a discussion of which we defer to Sec.~\ref{sec:sstop}.

\begin{figure*}[htbp]
\center
\includegraphics[width=.5 \textwidth]{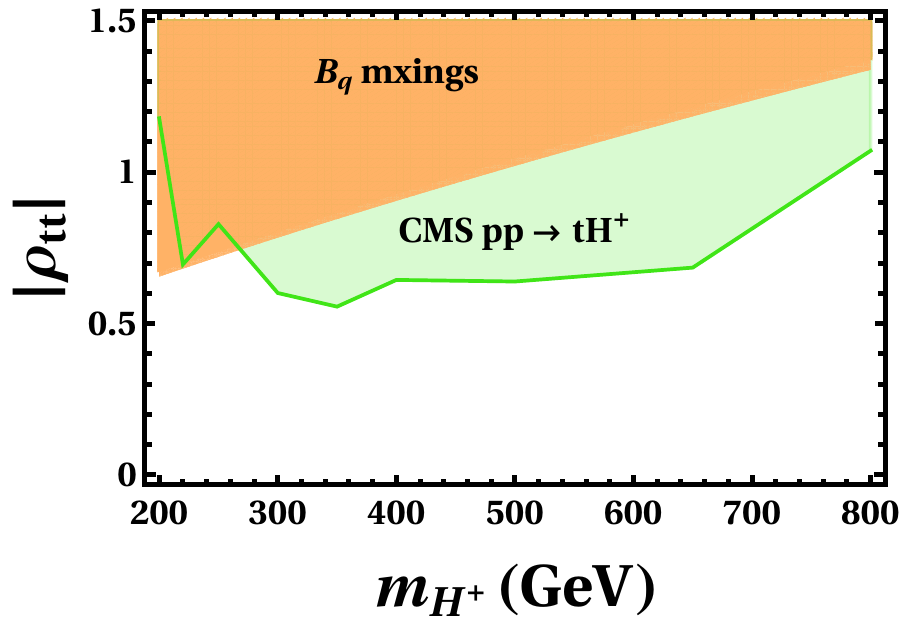}
\caption{
Exclusion regions of $\rho_{tt}$ from $B_{s,d}$ mixings (orange) and CMS search of
$pp\to \bar t {H}^+ (b) \to t \bar t \bar b (b)$~\cite{CMS:2020imj} (green) in the $|\rho_{tt}|$--$m_{H^+}$ plane.
}
 \label{rttconst}
\end{figure*}

The $\rho_{tt}$ is also constrained by indirect and direct measurements. 
E.g. the $t\bar t h$ coupling measurement by ATLAS~\cite{ATLAS:2018mme} and CMS~\cite{CMS:2018uxb} can constrain $\rho_{tt}$ but weak~\cite{Hou:2018uvr}.
This is primarily due to the suppression from the mixing angle $c_\gamma$, as can be seen from Eq.~\eqref{eff}. 
However, the $B_{s,d}$ mixing amplitude, where $\rho_{tt}$ enters at one loop through 
$tbH^\pm$ vertex, provides by far the most stringent limit with milder 
constraint also appears from $B\to X_s\gamma$ ($\mathcal{B}(B\to X_s\gamma)$) measurement.
We plot the exclusion region for $|\rho_{tt}|$ from $B_{s,d}$ mixing by allowing $2\sigma$ error on the UTfit values~\cite{UTfitBsmix} 
in Fig.~\ref{rttconst}. The direct search limits from  $pp\to \bar t {H}^+ (b) \to t \bar t \bar b (b)$~\cite{CMS:2020imj,ATLAS:2020jqj}
are stronger than $B_{s,d}$ mixings~\cite{Ghosh:2019exx} specially for $m_{H^\pm}\gtrsim 250$ GeV. We show the stronger 
CMS limit~\cite{CMS:2020imj}  by green shaded region in Fig.~\ref{rttconst} assuming $\mathcal{B}({H}^\pm\to t b)=100\%$.
The limits from $pp\to H/A \to t \bar t$~\cite{ATLAS:2017snw,CMS:2019pzc} and $pp\to t \bar t H/A \to t \bar t t \bar t$~\cite{CMS:2019rvj}
are somewhat weaker than those from $B_{s,d}$ mixings or $m_{H^\pm}$ search~\cite{Ghosh:2019exx}.

%%%%%%%%%%%%%%%%%%%%%%%%%%%%%%%%%%%%%%%%%%%%%%%%%%%%%%%%%%%
\section{Charged Higgs induced single-top production}\label{sec:bhpm}
%%%%%%%%%%%%%%%%%%%%%%%%%%%%%%%%%%%%%%%%%%%%%%%%%%%%%%%%%%%
First we discuss the discovery potential of $cg\to b H^+$ followed by $H^+ \to t \bar b$ decay~\cite{Ghosh:2019exx}.
The process is induced if $\rho_{tc}$ and $\rho_{tt}$ are non vanishing with both production and decay are $V_{tb}$
proportional. At the LHC it can be searched via $pp \to b H^+ + X \to b t \bar b +X$, with subsequent decay of $t\to b \ell \nu_\ell$,
constituting three $b$-jets, one charged lepton ($\ell =e,\;\mu$), and missing transverse energy $E^{\mbox{miss}}_T$ 
signature. Note that $\rho_{tc}$ and $\rho_{tt}$ coupling will also induce PDF suppressed 
$bg\to \bar c H^+ \to \bar c t \bar b$ and, $bg\to \bar t H^+ \to \bar t c \bar b$ which will contribute to the 
overall signature if the $c$-jet is misidentified as $b$-jet. 
Furthermore, $\rho_{tt}$ can induce subdominant $bg\to \bar t H^+ \to \bar t t \bar b$ process, which, 
with one top quark decaying semileptonically and other hadronically would also 
contribute to the same final state topology. We have accounted for all these contributions 
together in our analysis. We remark here that $bg\to \bar t H^+ \to \bar t t \bar b$ 
is the usual search mode for $H^\pm$ discovery for the ATLAS~\cite{ATLAS:2018ntn,ATLAS:2021upq} 
and CMS~\cite{CMS:2019rlz,CMS:2020imj} which is PDF suppressed in contrast to $cg\to b H^+ \to b t \bar b$.

The dominant SM backgrounds are QCD  
$t\bar t+$jets production and $s$- and $t$-channel single-top ($tj$), $Wt$ productions.
The subdominant contributions also arise from $t\bar t h$ and, $t\bar t Z$ with negligible fractions
come from $Z/\gamma^*$+jets, four-top ($4t$), inclusive $W$ production, $t\bar t W$, $tWh$. We combine these 
subdominant and negligible contributions together and denote as ``Others''.

We generate signal and background events at the leading order (LO)
in Monte Carlo event generator MadGraph5\_aMC@NLO~\cite{Alwall:2014hca} (denoted as MadGraph5\_aMC) 
with the parton distribution function (PDF) set NN23LO1~\cite{Ball:2013hta} at $\sqrt{s}=14$ TeV.
The generated events are then interfaced with 
PYTHIA~\cite{Sjostrand:2006za,Sjostrand:2014zea} for showering and hadronization. To incorporate the 
detector effects we utilized fast detector simulator Delphes~3~\cite{deFavereau:2013fsa} on our analysis.
The the matrix elements (ME) for the all the dominant and most subdominant backgrounds 
are generated with at least one additional jet, followed by ME and  
parton shower merging via MLM matching scheme~\cite{Mangano:2006rw,Alwall:2007fs}.
The jets are reconstructed with the  anti-$k_t$ algorithm with radius parameter $R=0.5$. 
The effective model is implemented is FeynRules~\cite{Alloul:2013bka}.

The LO cross section of the $t\bar t +$jets background is normalized up to the
NNLO by a factor $1.84$~\cite{twiki}. The $s$- and $t$-channel  single-top productions are accounted 
up to the NNLO cross sections by factors 1.47 and 1.2 respectively~\cite{twikisingtop} while
$Wt+$jets cross section is adjusted up to the NLO by factor 1.35~\cite{Kidonakis:2010ux}.
Further, the subdominant $t\bar t Z$ and $t\bar t h$ cross sections are corrected to the corresponding NLO 
ones respectively by factors 1.56~\cite{Campbell:2013yla} and 1.27~\cite{twikittbarh}.
The $Z/\gamma^*$+jets background is normalized up to  NNLO cross sections by factor 1.27~\cite{Li:2012wna},
whereas the tiny $4t$ and $t\bar t W$ backgrounds are up to NLO by factors
2.04~\cite{Alwall:2014hca} and 1.35~\cite{Campbell:2012dh} respectively.
The backgrounds $tWh$ and $W+$jets are kept at the LO.
For simplicity the charge conjugate processes assumed to have the same correction factor throughout the paper
and the signal cross sections are kept at LO. We take two benchmark masses $m_{H^\pm}=300$ and 500 GeV for 
with $\rho_{tc}=0.4$ and $\rho_{tt}=0.6$ for illustration.

\begin{table}[hbt!]
\centering
\begin{tabular}{|c |c| c| c| c | c| c| c |c |c|}
\hline
&&&&&&&\\ 
  Benchmark          & $t\bar t+$  & Single-top      & $Wt+$     &  $t\bar t h$   &  $t\bar t Z$    & Others   & \ Total  \          \\
  masses                &  jets       &                 &jets       &                &                 &          &   Bkg.  \\      
&&&&&&& (fb)\\                                
\hline
\hline
&&&&&&&\\
     $m_{H^\pm} = 300$ GeV &  1348.02   & 41.31            & 33.2      & 4.24           & 1.52           & 3.14      & 1431.43  \\
     $m_{H^\pm} = 500$ GeV  &  884.64    & 25               & 20.78     & 2.88           & 1.03           & 1.85       & 936.13   \\
 
\hline
\hline
\end{tabular}
\caption{The background cross sections (in fb) of the different backgrounds of the $pp \to b H^+ + X \to b t \bar b +X$
process after selection cuts at $\sqrt{s}=14$ TeV~\cite{Ghosh:2019exx}.}
\label{bkgcomptbhpm}
\end{table}

\begin{table}[hbt!]
\centering
\begin{tabular}{|c |c| c| c | c }
\hline
&&\\ 
Benchmark       &  \ Signal \         &  \ Significance ($\mathcal{Z}$)     \\ 
masses                               &       (fb)          &    300 (600) fb$^{-1}$    \\      
&& \\                                
\hline
\hline
$m_{H^\pm} = 300$ GeV       & 12.5                  &  5.7 (8.1)                                     \\ 
$m_{H^\pm} = 500$ GeV      & 10.01                 &  5.6 (8)                                      \\
\hline
\hline
\end{tabular}
\caption{The signal cross sections of the \bhpm process for $m_{H^\pm}=300$ and 500 GeV. The 
respective significances with 300 (600) \fbi integrated luminosities are given in third column~\cite{Ghosh:2019exx}. }
\label{sigtbhpm}
\end{table}

To reduce the backgrounds following event selection cuts are applied. The events are selected 
such that they contain at least one lepton, at least three $b$-jets and
some $E^{\mbox{miss}}_T$. The minimum transverse momentum ($p_T$) of the leading lepton is required to be $> 30$ GeV 
while for all three $b$-jets $p_T>20$ GeV. The maximum pseudo-rapidity ($|\eta|$) for the 
of all three $b$-jets and the lepton should be $<2.5$. 
The minimum separation ($\Delta R$) between any $b$-jet and lepton  
and, between any two $b$-jets are needed to be $> 0.4$. In each event $E^{\mbox{miss}}_T$ is required to be $ >35$ GeV.
Finally, $H_T$, defined as the sum of $p_T$ of the leading three $b$-jets and the leading lepton, 
should be $>350$ GeV and $>400$ GeV respectively for $m_{H^\pm}=300$ and 500 GeV respectively.

We compute the significance using the likelihood formula for a simple counting experiment as~\cite{Cowan:2010js}
\begin{align}
  Z(n|n_\text{pred})= \sqrt{-2\ln\frac{L(n|n_\text{pred})}{L(n|n)}},
  \qquad \text{with} \qquad L(n|\bar{n}) = \frac{e^{-\bar{n}} \bar{n}^n}{n !} \, \label{poisso},
\end{align}
where $n$ is the number of observed events and $n_\text{pred}$ predicted events.
For discovery, we perform hypothesis test between the observed signal plus background to the
background prediction with $Z(s+b|b) > 5$ is required for discovery. For exclusion limits we demand $Z(b|s+b) > 2$.

Let us discuss Table~\ref{sigtbhpm} and the discovery potential of \bhpm. 
We found that, the achievable statistical significance for $m_{H^\pm}=300$ GeV is
$\sim 5.7\sigma$ and for $m_{H^\pm}=500$ GeV $\sim 5.6\sigma$ at 300 \fbi integrated luminosity.
This illustrates that full Run-3 data can discover the single-top signature while
collected Run-2 data ($\sim 137$ \fbi) could reach to $3.9\sigma$ and $3.8\sigma$ significances.
%%%%%%%%%%%%%%%%%%%%%%%%%%%%%%%%%%%%%%%%%%%%%%%%%%%%%%%%%%%
\section{Same-sign top}\label{sec:sstop}
%%%%%%%%%%%%%%%%%%%%%%%%%%%%%%%%%%%%%%%%%%%%%%%%%%%%%%%%%%%
The $\rho_{tc}$ coupling can induce $cg \to tA/tH$, followed by 
$A/H \to t\bar{c}, \bar t c $ in g2HDM. 
The $t\bar t c$ final state would be diminished by overwhelming QCD $t\bar t$ production, 
however, $tt\bar{c}$ followed by semileptonically decaying top quarks give exquisite 
same-sign dilepton signature. Such signature would provide an excellent probe for the $\rho_{tc}$ coupling at the LHC.
No dedicated same-sign search has been performed at the LHC however some existing searches of ATLAS and CMS can
provide meaningful constraints. We note that $\rho_{tc}$ coupling may also induce
subdominant  $cg \to t t \bar c$ and $t$-channel $H/A$ 
exchanged $cc\to tt$ processes  which we included in our analysis.

The search for SM $4t$ production by CMS~\cite{CMS:2019rvj} 
based on 137~fb$^{-1}$ data at $\sqrt{s}=13$ TeV can constrain the parameter space 
for same-sign top. The search has different signal (SRs) and control (CRs)
regions based on the number of charged leptons ($e$, $\mu$) and $b$-tagged jets~\cite{CMS:2019rvj}. 
It was shown that the or $t\bar t W$ control region (denoted by CRW)~\cite{CMS:2019rvj} provides 
the best constraints for $|\rho_{tc}|$~\cite{Hou:2018zmg,Hou:2019gpn,Hou:2020ciy,Hou:2020chc}.

As per the baseline selection criterion of Ref.~\cite{CMS:2019rvj}, each event should contain 
two same-sign leptons and two $b$-jets and missing transverse momenta. The leading and subleading leptons are required to have
$p_T > 25$ and $20$\;GeV respectively with muon and electron $|\eta| < 2.5$\;($2.4$). 
All jets should satisfy $|\eta| < 2.4$. In addition to these, the CRW region can have up to five jets with the rest of the selection
cuts goes as follows: Events are selected if the $p_T$ of the jets and $b$-jets fulfill 
any of the following three selection criteria~\cite{info-Jack}:
(i) both the $b$-jets should have $p_T$\,$>$\,$40$\;GeV;
(ii) one $b$-jet with $p_T$\,$>$\,$20$\;GeV and $20$\,$<$\,$p_T$\,$<$\,$40$\;GeV for the second $b$-jet, but $p_T$\,$>$\,$40$\;GeV for the third jet;
(iii) both $b$-jets should satisfy $20$\,$<$\,$p_T$\,$<$\,$40$\;GeV,
but with two extra jets each with $p_T$\,$>$\,$40$\;GeV.
The scalar sum of the $p_T$ of all jets ($H_T$) is required to be $H_T$\,$>$\,$300$\;GeV and $p_T^{\rm miss}$\,$>$\,$ 50$\;GeV.
To reduce the charge-misidentified Drell-Yan ($Z/\gamma^*$) background CMS vetoed events with same-sign electron pairs 
lying within $m_{ee}$\,$<$\,$12$\;GeV. With the above selection criteria $335 \pm 18$ events were expected
(SM backgrounds plus $4t$) in the CRW while 338 events have been observed~\cite{CMS:2019rvj}.

\begin{figure*}[htbp]
\center
\includegraphics[width=.48 \textwidth]{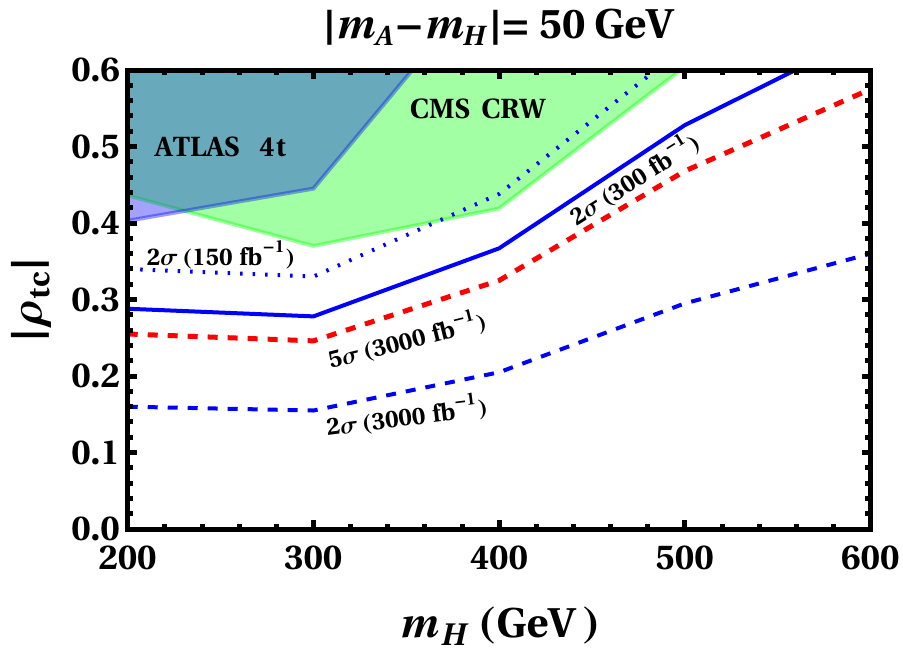}
\includegraphics[width=.48 \textwidth]{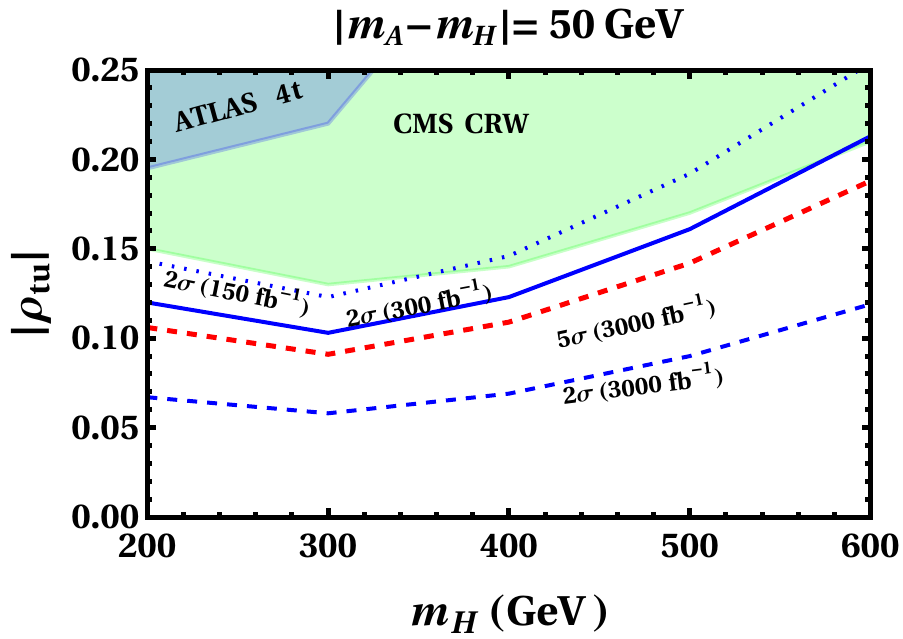}
\caption{
Exclusion limits (blue) and discovery reaches (red) 
for $|\rho_{tc}|$ and $|\rho_{tu}|$ by the same-sign top signature for various integrated luminosities at the 14 TeV LHC.
The purple and cyan regions are excluded respectively by CMS CRW~\cite{CMS:2019rvj}
and ATLAS CRttW2$\ell$~\cite{ATLAS:2020hrf} control regions.
}
 \label{discexclu}
\end{figure*}

For sizable $\rho_{tc}$ the $cg \to tA/tH \to t t \bar c$ processes abundantly populate the CRW region 
leading to stringent constraints on $\rho_{tc}$. However, if $H,\,A$ are  nearly degenerate in mass and width 
strong cancellation between the $cg \to tA \to tt\bar c$ and $cg \to tH \to tt\bar c$ happen. The cancellation
is exact if both mass and widths are degenerate~\cite{Kohda:2017fkn}. This can be simply understood
from Eq.\ref{eff}, where the $cg \to tA \to t t \bar c$ amplitude has an additional a
factor of $i^2 = -1$ to that $cg \to tH \to t t \bar c$. In order to avoid such cancellation 
and break the mass and width degeneracy we take $|m_A-m_H|=50$ GeV. Assuming  $\rho_{tc}$ is only non-vanishing coupling
and  demanding the sum of the $\rho_{tc}$-induced events and expected SM 
and agree with observed number of events within the $2\sigma$ error bars we show the excluded region from CRW
in Fig.~\ref{discexclu} (left) green shaded regions. We also provide the constraint on 
$\rho_{tu}$ for comparison. Here we assumed Gaussian behavior for simplicity.

A similar $4t$ search has been performed by ATLAS~\cite{Aad:2020klt}
with full Run-2 data but the selection criteria categorizing into different CRs and SRs are somewhat 
different from that of CMS. It has been found that the  $t\bar t W$, called $\rm{CRttW2\ell}$ of Ref.~\cite{Aad:2020klt} is the
most relevant one. The CR is defined to have least two same-sign leptons ($e\mu$ or $\mu\mu$),
at least four jets with at least two of them being $b$-tagged. For the same-sign leptons 
$p_T > 28$\;GeV with $|\eta^\mu| < 2.5$ and $|\eta^e| < 1.5$ whereas 
all the jets should satisfy $p_T > 25$ GeV and $|\eta| < 2.5$.
The scalar $p_T$ sum over all jets and same-sign leptons (different from CMS), $H_T < 500$\;GeV. 

Unlike CRW for CMS, ATLAS does provide the expected and observed events in $\rm{CRttW2\ell}$, however,
provide a figure of comparison between prediction and data for the variable $\sum p_T^\ell$ (see Ref.~\cite{Aad:2020klt} for detailed definition).
Following Ref.~\cite{Bonilla:2017lsq} prescription of digitizing the figure, Ref.~\cite{Hou:2020ciy}
found the number of expected and observed events in the $\rm{CRttW2\ell}$ from this distribution to be
$378\pm 10$ and $380$ respectively. Here errors for the expected events in each bin of the $\sum p_T^\ell$ distributions 
are simply added in quadrature. The exclusion region from $\rm{CRttW2\ell}$ is shown in Fig.~\ref{discexclu}
by blue shaded regions. Here we followed the same procedure as in before for event generation, hadronization, showering while 
utilized CMS based detector card of Delphes. At this point we remark that he ATLAS collaboration has
made several similar searches~\cite{ATLAS:2019fag,ATLAS:2018alq}
constraints are much weaker due too strong selection cuts~\cite{Hou:2020chc}.

% ------------------------------------------------------------------
\subsection{A dedicated same-sign top search}
% ------------------------------------------------------------------
Although the CMS and ATLAS $4t$ search with Run~2 data can provide meaningful constraints, however,
the searches are not optimized for $cg \to tA/tH \to t t \bar c$ processes. 
Here we discuss a dedicated same-sign top search via $pp\to tA/tH +X \to t t \bar c +X$ followed 
by semileptonic decay of both the top quarks with two same-sign dileptons ($e$, $\mu$), three jets with at least 
two $b$-tagged jets and missing energy. We denote this signature as SS$2l$-$2bj$.

There are several SM backgrounds for SS$2l$-$2bj$ topology. 
The dominant ones are  $t\bar t W$, $t\bar t Z$, $4t$, $tZ +$\,jets and $t\bar t h$. 
Further if a lepton charge is misidentified (denoted as charge- or $Q$-flip) 
the $t\bar t+$\,jets and $Z/\gamma^*+$\,jets processes would also contribute to the 
overall background cross section with a misidentification efficiency
$2.2\times 10^{-4}$~\cite{ATLAS:2016kjm,Alvarez:2016nrz,Aaboud:2018xpj}.
A CMS study~\cite{CMS:2017tec} with similar final state topology, albeit slightly different 
selection cuts, reported the ``nonprompt'' and fake backgrounds are that of $\sim 1.5$ times 
the $t\bar{t}W$ background. As the nonprompt and fake backgrounds are not ideally
modeled in Monte Carlo simulations, for simplicity we just add this component to the overall background 
at 1.5 times that of $t\bar t W$ background after our selection cuts.
There of course are some tiny backgrounds, such as $3t + W$ and $3t + j$ etc. which we have neglected
in our analysis. For event generation, hadronization, showering and detector effects
we follow the same simulation chains as in previous section. Moreover we utilize 
the same $K$ factors for different backgrounds as described earlier while keep the signal cross sections
at LO.

We the adopt following selection cuts for the SS$2l$-$2bj$ search~\cite{Hou:2020ciy,Hou:2020chc}.
The leading and subleading leptons are need to have $p_T > 25$ and 20\,GeV respectively, while pseudo rapidity should be
$|\eta| < 2.5$. The jets are required to have $p_T> 20$\;GeV and $|\eta| < 2.5$ with $E^{\rm miss}_{T} > 35$  GeV.
The $\Delta R$ between any two visible objects should satisfy $\Delta R > 0.4$. 
Finally, we apply scalar sum of the $p_T$ of the two same-sign leptons and two leading $b$-tagged jets 
and the leading non $b$-tagged jets should be $ > 300$\;GeV. The background cross sections with the above selection cuts
are provided in Table~\ref{backg}. We estimate the signal cross sections for a
reference $\rho_{tc}=1$ for few discrete values of $m_H$  with $|m_A-m_H|=50$ GeV and assuming $m_H > m_A$
with $\mathcal{B}(A/H\to t \bar c + \bar t c) = 100\%$. We then scale the signal cross section simply 
with $|\rho_{tc}|^2\mathcal{B}(A/H\to t \bar c + \bar t c)$ and provide  discovery (red) and exclusion (blue) contours
for in Fig.~\ref{discexclu}. A similar procedure is repeated for $\rho_{tu}$ induced SS$2l$-$2bj$ for  
the right panel of Fig.~\ref{discexclu}.
For a detail discussion on SS$2l$-$2bj$ we redirect readers to 
Refs.~\cite{Hou:2020ciy,Hou:2020chc}.

\begin{table}[t!]
\centering
\begin{tabular}{c |l c c c }
\hline

                     \,Backgrounds\,                 & \ \,Cross section (fb)\,     
& \\
\hline
\hline
&\\
                       $t\bar{t}W$               & \hskip1.1cm 1.31               \\
                       $t\bar{t}Z$                & \hskip1.1cm 1.97               \\
                       $4t$                           & \hskip1.1cm 0.316               \\
                       $tZ+$\,jets                 & \hskip1.1cm  0.255                \\
                      $t\bar t h$                   & \hskip1.1cm  0.07                \\                       
                     $Q$-flip                       & \hskip1.1cm  0.024                 \\
                     nonprompt                      & \hskip0.8cm $1.5\times t\bar{t}W$      \\
\hline
\hline
\end{tabular}
\caption{The background cross sections for a dedicated SS$2l$-$2bj$ search.
}
\label{backg}
\end{table}

Let us understand the Fig.~\ref{discexclu}. A dedicated SS$2l$-$2bj$ search could 
exclude $|\rho_{tc}|\sim 0.5$ for $m_H\sim 200$--400 GeV with $|\rho_{tc}|\sim 0.6$--0.7 for  $m_H\sim 500$--700 GeV.
More stringent limit is achievable with full Run-3 data while
high luminosity LHC (HL-LHC) i.e. with full 3000 fb$^{-1}$ integrated luminosity
one $|\rho_{tc}|\sim 0.1$ can be excluded for $m_H\sim 200$--400 GeV whereas reaching up to
$|\rho_{tc}|\sim 0.2$ for $m_H\sim 500$--700 GeV. A sensitive is possible for 
$\rho_{tu}$ induced same-sign top signature, as can be seen from left panel of Fig.~\ref{discexclu}.
Note here we set all other couplings to zero except for $\rho_{tc}$ which simply leads 
same decay widths for $A$ and $H$.  If the mass difference $|m_A-m_H|$ is reduced
the $cg\to tA \to t t \bar c$ and $cg\to tH \to t t \bar c$ amplitudes will cancel each other
making the same-sign top signature diminished.

%%%%%%%%%%%%%%%%%%%%%%%%%%%%%%%%%%%%%%%%%%%%%%%%%%%%%%%%%%%
\section{Triple top}\label{sec:3top}
%%%%%%%%%%%%%%%%%%%%%%%%%%%%%%%%%%%%%%%%%%%%%%%%%%%%%%%%%%%  
The triple-top signature $cg\to tA/tH \to t t  \bar t$ is induced 
if both $\rho_{tc}$ and $\rho_{tt}$ are nonzero. This process 
can be searched at the LHC via $pp\to tA/tH +X \to t t  \bar t +X$
with at least three leptons ($\ell=$ $e$, $\mu$) and three $b$-jets 
and $E^{\rm miss}_T$. We remark that unlike the same-sign top signature 
the $pp\to tA +X \to t t  \bar t +X$ and $pp\to tH +X \to t t  \bar t +X$
processes do not cancel each other when both masses and decay widths  
are degenerate, the consequence of which we shall discuss shortly.
This is because for a real $\rho_{tt}$, the $cg\to tA \to t t  \bar t$ amplitude gains
a factor of $i^2\gamma_5$ to that of the $cg\to tH \to t t  \bar t$~\cite{Kohda:2017fkn}.

The dominant SM backgrounds as in before are $t\bar t Z+$jets, $t\bar t W+$jets,
$4t$ with subdominant contributions come from $t\bar t h$, $tZ$+jets, $t\bar t h$, $3t +j$ and $3t + W$.
The overwhelming $t\bar t +$jets process also contributes to the overall
background contribution if one of the jet gets misidentified as a charged lepton.
As discussed above such backgrounds are not properly modeled in Monte Carlo event generator such 
as MadGraph5\_aMC. Therefore we estimate such contribution by simply 
applying the selection cuts on the $t\bar t +$jets samples with one of the jet is taken 
as the third lepton with misidentification efficiency $\epsilon_{\rm fake} = 10^{-4}$~\cite{ATLAS:2016dlg,Alvarez:2016nrz}.
The QCD $K$ factors for different backgrounds are same as in previous sections with 
$3t +j$ and $3t + W$ and signal are kept at LO for simplicity.

In order to distinguish the signal events from the backgrounds 
following event selection cuts are applied. The events are selected 
such it contain at least three leptons, at least $b$-jets three
and $E^{\mbox{miss}}_T$. The three leading $b$-jets and leptons are required to have
$p^{\ell}_{T} > 20$ GeV and $p^{b}_T > 20$ GeV respectively and $E_T^{\rm miss} > 30$ GeV.
The pseudo-rapidity .
We further use the scalar sum of transverse momenta of the 
three leading $b$-jets and three leading leptons should be $> 300$ GeV.
Finally, to reduce contributions from $t\bar t Z +$ and $tZ$-jets backgrounds,
we reject events when a same flavor, opposite charged lepton pair lie within 
$80~\mbox{GeV}< m_{\ell\ell} < 100~\mbox{GeV}$ window. For events, if more than one such pair
exist the veto is applied to the pair to $m_Z$.

\begin{table*}[htbp!]
\centering
\begin{tabular}{| c | c |c| c| c| c | c| c| c |c |c|}
\hline
&&&&&\\ 
$t\bar{t}Z$  &  $4t$   &  $t\bar{t}W$   & Others & $t\bar t$+jets & \ Total  \ Bkg.  (fb)       \\                                
\hline
\hline

           &       &   &      &  &  \\
 0.135     & 0.105     & 0.011  & 0.007 & 0.1660   & 0.424 \\

\hline
\end{tabular}
\caption{Different background cross sections for triple-top signature. In the last column 
the total background cross sections is presented.}
\label{triple-top-bkg}
\end{table*}

\begin{table}[hbt!]
\centering
\begin{tabular}{|c |c| c| c | c }
\hline
&&\\ 
 Benchmark&  \ Signal \         &  \ Significance ($\mathcal{Z}$)     \\ 
masses                              &       (fb)          &    600 (3000) fb$^{-1}$    \\      
&& \\                                
\hline
\hline
$m_H = m_A$ = 400 GeV       & 0.075                  &  2.7 (6.1)                                     \\ 
$m_H = m_A$ = 500 GeV       & 0.063                  &  2.3 (5.2)                                      \\
\hline
\hline
\end{tabular}
\caption{The signal cross sections of the triple-top process for two different benchmark masses. The 
respective significances with 600 (3000) \fbi integrated luminosities are given in third column. }
\label{triple-top-sig}
\end{table}

The signal and background cross sections are presented in Table~\ref{triple-top-sig} and 
Table~\ref{triple-top-bkg}. It is clear that triple top signature can be discovered at the 
HL-LHC provided $\rho_{tc}\sim 1 $ and $\rho_{tt} \sim 1$ while evidence might come 
with 600 \fbi. It may seem that $\rho_{tc}\sim 1$ is excluded from the left panel Fig.~\ref{discexclu},
,however, the exclusion region was found assuming all other couplings are vanishingly small. However,
it is shown that an $\mathcal{O}(1)$ is still allowed if $\rho_{tt} \sim 1$, where the latter coupling 
will suppress the same-sign top cross section via $A/H\to t \bar t$ decay~\cite{Hou:2019gpn}.

We now remark on the consequence of the discovery of the different heavy Higgs induced multi-top productions 
in the light of EWBG and inflation. It is found that complex $10^{-2}\lesssim\rho_{tt} \gtrsim 1$ and $\rho_{tc}\gtrsim 0.5$~\cite{Fuyuto:2017ewj}
individually can account for the observed baryon asymmetry of the Universe. Therefore, it is clear that $cg\to b H^+ \to b t \bar b$, 
$cg\to tA/tH\to tt\bar c$ and $cg\to tA/tH\to tt\bar t$ all could provide potential discovery mode for $\rho_{tc}$ and $\rho_{tt}$ EWBG mechanism.
If inflation is realized in g2HDM it would require a nearly mass spectra for $m_A$, $m_H$, $m_{H^\pm}$~\cite{Lee:2021rzy,Modak:2020fij}. With nonvanishing $\rho_{tc}$ and $\rho_{tt}$,
for such a parameter space, the same $cg\to b H^+ \to b t \bar b$ signature would emerge first but the same-sign top signature could be diminished significantly
due to cancellation for near degeneracy of masses and widths~\cite{Lee:2021rzy,Modak:2020fij}. On the other hand the triple-top signature 
will also emerge providing a smoking gun signature for cosmic inflation.

%%%%%%%%%%%%%%%%%%%%%%%%%%%%%%%%%%%%%%%%%%%%%%%%%%%%%%%%%%%%%%%%%%%%
\section{Summary and Outlook}\label{disc}
%%%%%%%%%%%%%%%%%%%%%%%%%%%%%%%%%%%%%%%%%%%%%%%%%%%%%%%%%%%%%%%%%%%%
We have discussed the discovery prospect of heavy Higgs induced single-top, same-sign, and triple-top 
productions in g2HDM. We have shown that for sizable $\rho_{tc}$ and $\rho_{tt}$ the $cg\to b H^+ \to b t \bar b$
and $cg\to tA/tH\to tt\bar c$ processes could be discovered with Run-3 data but $cg\to tA/tH\to tt\bar t$ may 
require HL-LHC data. We have shown that the $cg\to b H^+ \to b t \bar b$ could be discovered in the 
Run-3 if $m_{H^\pm}\sim 200$--$500$ GeV. Existing ATLAS and CMS data provide some constraint but a dedicated 
search for the $cg\to tA/tH\to tt\bar c$ should be performed. The Run-3 data may already provide sensitive probe for the same-sign top process 
if $m_A$ and $m_H$ lie in the 200--800 GeV range. 

We have turned off all $\rho_{ij}$ couplings except for $\rho_{tt}$ and $\rho_{tc}$ for simplicity.
If $\rho_{ii}\sim \lambda_i$ i.e. $\rho_{bb}\sim \lambda_b$ and  $\rho_{\tau\tau}\sim \lambda_\tau$
different decay modes of $A$, $H$, $H^\pm$ would open up. This would dilute the signatures mildly as long as these couplings are small.
In general $\rho_{bb}\sim 0.2$ is still allowed~\cite{Modak:2018csw} while $\rho_{tu}\sim 0.1$ is possible~\cite{Hou:2020ciy} for the sub-TeV 
$A$, $H$, $H^\pm$. Presence of such large couplings may not only dilute the signatures significantly the parameter space could also 
be subjected to several flavor physics observables~\cite{Crivellin:2013wna}. This would require a detail analysis which we leave out for future.

The heavy Higgs induced single-top, same-sign top and triple-top signatures may provide indirect probes to baryogenesis and inflation. 
While both these early Universe phenomena are connected to sub-TeV mass range,however, the parameter space are mutually exclusive~\cite{Lee:2021rzy,Modak:2020fij}. 
It should be noted that the triple-top process is already covered in the search program of ATLAS collaboration~\cite{ATLAS:2022xpz}. We also highlight 
a dedicated webpage for the same-sign and triple-top processes for experimental studies~\cite{twikifcnh} and similar page is being created for the charged Higgs induced 
single-top process. If discovered these processes would not only confirm beyond the Standard Model physics but may shed light on the mechanisms behind 
observed baryon asymmetry or the cosmic inflation.

\appendix

\vskip0.2cm
\noindent{\bf Acknowledgments} \
TM is supported by postdoctoral fellowship of Universit{\"a}t Heidelberg.

%%%%%%%%%%%%%%%%%%%%%%%%%%%%%%%%%%%%%%%%%%%%%%%%%%%%%%%%%%%%%%%%%%%%%%%%%%%%%%%%%%%%%%%%%%

\end{document}